\documentclass{llncs}
\usepackage{times}
\usepackage{url} 
\usepackage{graphicx}
\usepackage{time}
\usepackage{ifthen}
\usepackage{amssymb}
\usepackage{xspace}
\usepackage{moreverb}
\usepackage{boxedminipage}
\usepackage{array}

\makeatother

\newcommand{\secref}[1]{Section~\ref{Sec:#1}}

\newcommand{\curl}[1]{\footnote{~\scriptsize\url{#1}}}

\newcommand{\jhotdraw}{\textsc{JHotDraw}\xspace}

\newcommand{\aspectj}{\textsc{AspectJ}\xspace}
\newcommand{\ajhd}{\textsc{AJhotDraw}\xspace}
\newcommand{\bettar}{\textsc{Bettar}\xspace}

\newboolean{showcomments}
\setboolean{showcomments}{true}
\ifthenelse{\boolean{showcomments}}
  {\newcommand{\nb}[2]{
  \fbox{\bfseries\sffamily\scriptsize#1}
    {\sf\small$\blacktriangleright$\textit{#2}$\blacktriangleleft$}
   }
  }
  {\newcommand{\nb}[2]{}
   
  }

 \def\now{{\def\Time{3}\def\Hour{4}\def\Minute{5}%
  \count\Time=\time\relax\ifnum\count\Time=0\count\Time=1440\fi%
  \count\Hour=\count\Time\relax\divide\count\Hour by 60\relax%
  \count\Minute=\count\Hour\relax\multiply\count\Minute by -60\relax%
  \advance\count\Minute by \count\Time\relax\the\count\Hour\relax:%
  \ifnum\count\Minute<10 0 \fi\the\count\Minute\relax}}

\title{%
A Systematic Aspect-Oriented Refactoring and Testing Strategy,
and its Application to JHotDraw}

\titlerunning{Aspect-Oriented Refactoring and Testing}

\author{
       Arie van Deursen\inst{1,}\inst{2}
\and
       Marius Marin\inst{2}
\and
       Leon Moonen\inst{2,}\inst{1}
}

\authorrunning{van Deursen et al}

\institute{
   Centrum voor Wiskunde en Informatica (CWI), The Netherlands
\and
   Software Evolution Research Lab, Delft Univ.\ of Technology, The Netherlands
\\[.5\baselineskip]
  \email{Arie.van.Deursen@cwi.nl},~~\email{A.M.Marin@ewi.tudelft.nl},\\
   and \email{Leon.Moonen@computer.org}
}

\begin{document}
\maketitle

%%%%%%%%%%%%%%%%%%%%%%%%%%%%%%%%%%%%%%%%%%%%%%%%%%%%%%%%%%%%%%%%%%%%%
% For proceedings:
%\thispagestyle{empty}
%\pagestyle{empty}
%
% For review
\pagestyle{plain}
%%%%%%%%%%%%%%%%%%%%%%%%%%%%%%%%%%%%%%%%%%%%%%%%%%%%%%%%%%%%%%%%%%%%%

\begin{abstract} 
Aspect oriented programming aims at achieving
better modularization for a system's crosscutting concerns in order
to improve its key quality attributes, such as evolvability and
reusability. Consequently, the adoption of aspect-oriented techniques
in existing (legacy) software systems is of interest to remediate
software aging.  The refactoring of existing systems to employ
aspect-orientation will be considerably eased by a systematic approach
that will ensure a safe and consistent migration.

In this paper, we propose a refactoring and testing strategy that
supports such an approach and consider issues of behavior conservation
and (incremental) integration of the aspect-oriented solution with the
original system.  The strategy is applied to the \jhotdraw{} open
source project and illustrated on a group of selected concerns.
Finally, we abstract from the case study and present a number of
generic refactorings which contribute to an incremental
aspect-oriented refactoring process and associate particular types of
crosscutting concerns to the model and features of the employed aspect
language.  The contributions of this paper are both in the area of
supporting migration towards aspect-oriented solutions and supporting
the development of aspect languages that are better suited for such
migrations.

\end{abstract}

%%% Local Variables: 
%%% mode: latex
%%% TeX-master: t
%%% TeX-master: t
%%% End: 

\section{Introduction}

Aspect-oriented software development is a programming paradigm
that addresses \emph{crosscutting concerns}: behavior of a software
system that is hard to decompose and isolate in existing paradigms (as
object orientation) and requires its implementation to be spread across
many different modules. Aspect-oriented software development aims to
overcome these limitations by capturing such crosscutting behavior in
a new modularization unit, the aspect, and offers (compile time) code
generation facilities to weave aspect code into the rest of the
system. Claimed benefits include improved evolvability and reusability
of (parts of) the software system~\cite{discuss-aop01,laddad2003}.

Addressing the aforementioned modularization limitations and the
resulting code scattering and tangling does not only pay off in the
development of new applications but it will also have major benefits
in existing software systems where these, and associated, problems
have become known as software aging~\cite{parnas-icse94} or software
entropy~\cite{leh80a,Leh80b,LPR98}.

The adoption of aspect-orientation in existing software requires
refactoring: code transformations that improve the internal structure
of a system while preserving its external behavior.  Existing work on
aspect-introducing-refactorings has mainly focused on presenting
aspect-oriented solutions to typical crosscutting problems, especially
in the context of design patterns, and showing that this results in a
better separation of concerns
\cite{DPAOP02,laddad2003,MPM04,Monteiro03}.  Also tool support for
aspect extensions of refactorings, such as \emph{method extraction}, has
been investigated~\cite{EttinVerb04}.

We argue, however, that widespread adoption of aspect-oriented
techniques in existing software systems is still hindered by a number
of open issues:
\begin{itemize}
\item Lack of a systematic approach to refactor legacy code to employ
  aspect-oriented solutions; 
\item Proper understanding of the testing challenges that rise from
  behavior preserving migration towards a new or extended language,
  such as the development of an aspect-oriented fault model and
  the definition of an explicit test adequacy criterion;
\item Suitability of aspect languages. Analysis of crosscutting
  concerns that were identified in various object-oriented systems in
  earlier work (\cite{FanIn04}) suggested that we might encounter
  difficulties when trying to refactor those concerns into aspects:
  the mechanisms offered by particular aspect languages, or the joint
  point model behind the language were not always sufficient to
  capture all types of concerns that were encountered;
\item Availability of aspect-oriented and non-aspect-oriented
  implementations of the same software system which can be used to
  show the evolution benefits of proposed solutions; and, more
  generally, 
\item An overall assessment of the benefits of aspect-oriented
  software development.
\end{itemize}
These issues are the principal motivation for the work described in
this paper. To address them, we propose a language and system
independent refactoring and testing strategy to adopt aspect-oriented
solutions in legacy code.  The strategy consists of a number of
systematic steps that guide the transformation, ensure conservation of
observable behavior and help one deal with the intricacies of
aspect-oriented migrations.

We demonstrate suitability of the proposed strategy in a case study
in which we migrate \jhotdraw, a (relatively) large and well-designed
open source Java application, to \ajhd, a corresponding
aspect-oriented version which is based on \aspectj{}, an aspect
language extending Java with crosscutting
functionality. 

Based on the difficulties that were encountered during our refactorings, we
reflect on the suitability of particular (features of) aspect
languages for this type of work (i.e. evolution of legacy systems as
opposed to greenfield development). This  provides designers
of aspect languages with valuable insights into situations that
require model or feature extensions to address these concerns.

The remainder of the paper is structured as follows: In the next
section we propose an aspect-oriented refactoring and testing strategy
together with its accompanying fault model and test adequacy
criterion. This is followed by a section that presents general
considerations about the case study. Next, a number of selected
crosscutting concerns are discussed on an individual basis, depicting
the context in which they occur, the original and refactored
implementation together with the benefits and drawbacks. In
\secref{catalogrefacts}, a number of generic refactorings are
abstracted from the case study and associations are made between types
of crosscutting functionality and the aspect language model and
features.  We conclude with a general discussion followed by an
overview of related work, summary of our contributions and present
some directions for future work.

%%% Local Variables: 
%%% mode: latex
%%% TeX-master: t
%%% End: 

\section{The \bettar Refactoring Strategy}
\label{Sec:refactstrategy}

The refactoring strategy we propose is called \bettar:
its objective is to obtain
\underline{B}etter \underline{E}volvability \underline{T}hrough
\underline{T}ested \underline{A}spect \underline{R}efactorings.
We distinguish the following steps:

\begin{description}
  
\item[Identification of Crosscutting Concerns:] Search for candidate
  aspects using ``aspect mining'' techniques such as fan-in analysis
  or clone detection \cite{BDET04,FanIn04,TonCec04}.  Assess the
  scattering and tangling implications of the current
  non-aspect-oriented solution to the crosscutting concerns
  identified.
  
\item[Aspect Design:] Identify how the concern could be implemented as
  an aspect.  Assess the pros and cons and compare these with the
  existing solution.
  
\item[Refactoring Design:] Devise a sequence of (small) steps,
  refactoring the object solution to the aspect solution.  This may
  involve various traditional object-oriented refactorings, in order
  to \emph{unplug} the crosscutting concern from code implementing
  other concerns, in addition to refactorings moving functionality to
  aspects.

  Conduct trade-off analysis to determine whether the aspect benefits
  outweigh the refactoring costs.
  
\item[Test Suite Design:] Conduct a baseline test on the existing
  implementation, and analyze the test adequacy of the current test
  suite with respect to the risks introduced by the aspect-oriented
  solution as well as by the refactoring process itself.  If
  necessary, create or extend the test suite -- see the next section
  for further details on this step.
  
\item[Execute and Test:] Carry out the refactorings, and verify that
  the behavior of the system is unaltered by means of the test suite.

\end{description}

In Section~\ref{Sec:selectedrefacts} we will apply these steps to various cross
cutting concerns as occurring in the open source \jhotdraw system.
As the interplay between refactoring and testing, in particular in 
an aspect-oriented setting, has received very little attention in the literature
so far, we start by elaborating the testing steps.

%%% Local Variables: 
%%% mode: latex
%%% TeX-master: t
%%% End: 

\subsection{Ensuring Behavior Preservation while Refactoring to Aspects}
\label{Sec:teststrategy}

Refactoring is the process of changing a software system in such 
a way that it improves the code's internal structure,
\emph{without altering its external behavior} \cite{fowler1999}.
In order to ensure the latter constraint, most literature on
refactoring assumes the presence of a test suite that verifies
the correct functional behavior of the system to be refactored.
As long as this test suite is executed before and after each refactoring,
we can assume that we will be warned as soon as one of our
refactorings affects the correct behavior of the system.

In practice, however, the creation of such a test suite is challenged
by a number of issues. These hold in the general (pure
object-oriented) situation, as well as in a setting where the
refactoring includes the introduction of aspects. These issues are:
\begin{itemize}
\item To test effectively, testing should be based on a \emph{fault
    model}.  Such a fault model guides our search for test cases that
  give the highest probability of finding typical faults
  \cite{Binder:2000}.  The adoption of aspects opens opportunities for
  different types of faults, calling for an explicit aspect-oriented
  fault model.
\item Systematic testing makes use of an explicit test adequacy
  criterion (see, e.g., \cite{Binder:2000}), usually expressed as a
  coverage percentage to be achieved in some coverage model.
  Refactoring changes the internal structure of the code.  Since test
  adequacy is expressed in terms of code structures covered, a
  refactoring may very well affect coverage negatively --- a
  phenomenon referred to as the \emph{antiextensionality axiom} by
  Weyuker \cite{Weyuker:1988}.

\item In addition to new faults introduced by using aspects, the
  \emph{mechanics} of actually carrying out a refactoring may lead to
  a new fault.  For example, when we moved part of a method from a
  class to an aspect, we did not copy-paste all statements.
  At other times, refactorings affect the public interface of classes,
  for example when moving a public method.  This implies that the
  \emph{test suite} needs to be adapted as well, causing an extra risk
  of letting errors pass \cite{DM2002.VSR}.
\end{itemize}

This calls for a testing approach that is dedicated to refactorings
involving the introduction of aspects.  In this section, we provide
such an approach.  In order to do so, we first propose an
aspect-oriented fault model, as well as aspect-oriented test adequacy
criteria.  Note that such a model and criteria can never be complete:
we believe, however, that our proposals represent an important first
step.

The specific faults that can be made while refactoring depend on the
actual refactoring applied.  Therefore, we will not provide a general
fault model for refactorings, but will indicate typical faults and
testing implications when discussing some of the individual
refactorings that we have used. We will then also indicate whether the
refactoring may require changes to existing test cases.

The test strategy presented below is applicable to any development
project making use of aspects, and as such is independent of our case
study on \jhotdraw.  In the later sections we will discuss how we
actually applied the proposed fault model and adequacy criteria when
while refactoring \jhotdraw to \ajhd.

\subsection{An Aspect-Oriented Fault Model}

A \emph{fault model} identifies relationships and components of the
system under test that are most likely to have faults
\cite{Binder:2000}.  We distinguish faults for \emph{inter-type
  declarations}, \emph{pointcuts}, and \emph{advice}.

Inter-type declarations are most error-prone (and most powerful) when
used to create polymorphic functions.  Therefore, our fault model for
introductions is based on Binder's existing fault models for
polymorphism and inheritance~\cite[p.501]{Binder:2000}.  Our model
distinguishes the following faults that are specifically related to
\emph{polymorphism in introductions} and \emph{inter-type
  declarations}:
\begin{itemize}
\item Wrong method name in introduction, leading to a missing or
  unanticipated method override.
\item Wrong class name in a member-introduction, leading to a method
  body in the wrong place in the class hierarchy.
\item Inconsistent parent declaration, resulting in a (sub)class that
  violates Liskov's and Wing's behavioral notion of subtyping
  \cite{LW94} and/or Meyer's design-by-contract rules for inheritance
  (such as \emph{require no more, ensure no less}) \cite{Mey97}.
\item Inconsistent overridden method introduction, also resulting in a
  violation of behavioral subtyping.
\item Omitted parent interface resulting in a method that was intended
  to implement an interface method, but which now stands on its own.
\end{itemize}

Faults in \emph{pointcuts} will have the effect that advice code is
activated at the wrong program execution points.  Such faults include:
\begin{itemize}
\item Wrong primitive pointcut, using, for example, a \texttt{call}
  instead of an \texttt{execution} construct.
\item Errors in the conditional logic combining the 
  individual pointcut conditions.
\item Wrong type, method, field, or constructor pattern in pointcut.
  In particular, the use of \texttt{*} as a pattern wildcard or in
  string matching easily leads to too many join points.  Furthermore,
  if the underlying classes are modified or extended, the wildcard may
  become erroneous without the compiler being able to notice this.
\end{itemize}

Faults in \emph{advice} will result in the wrong action at a certain
point of execution. Such faults include:
\begin{itemize}
\item Wrong advice specification (using \texttt{before} instead of
  \texttt{after}, using \texttt{after} with the wrong argument, etc.).
\item Wrong or missing \texttt{proceed} in \texttt{around} advice.
\item Wrong or missing advice precedence.
\item Advice code causing a method to break its class invariant or to
  fail to meet its postcondition.
\end{itemize}

The fault model above states that aspect weaving should not conflict
with class invariants or method pre- and postconditions.  The safe
route to follow is that class resulting from weaving is a proper
subtype of the original class. Put in terms of design by contract, the
class invariant of the resulting class cannot be weaker, its method
preconditions cannot be stronger, and the postconditions cannot be
weaker.  Typical examples are ``harmless'' aspects which add logging
or tracing.  In this situation, existing code using the class need not
be aware that new functionality has been woven into it.  In other
words, the test suite for the original classes should pass on the
classes extended by introduction or advice as well, and doing so will
help to find faults originating from improper extensions.

An alternative route is that the aspect actually modifies the contract
in a way that conflicts with the inheritance rules from
design-by-contract.  This may include changes in method pre- or
postconditions, and may thus require weaving in additional code at all
affected call sites.  A typical example is an aspect that adds
security checks: this may lead to additional exceptions which at some
point should be handled in the original application.  Faults in this
approach will not be restricted to the newly woven class, but may be
at any call site in the application.  This setting is much harder to
test and immediate reuse of the test suite will not be possible.

Similar distinctions are made by Clifton and Leavens \cite{CL03}, who
discuss the relation between behavioral subtyping and aspect weaving,
and distinguish \emph{observers} from \emph{assistants}.  Rinard
\emph{et al} \cite{RSB04} classifies interactions between woven code
and the original code, recognizing \emph{augmentation},
\emph{narrowing}, \emph{replacement}, and combinations between them.
The key concern of these authors is modular reasoning: in our setting
it is modular testing, and reuse of test suites to woven classes.

\subsection{Aspect-Oriented Test Adequacy}

A \emph{test adequacy criterion} prescribes the elements of the
implementation under test that need to be exercised by a test suite.
The \emph{coverage} achieved by a test suite is the percentage of
elements actually exercised.  In this section we formulate adequacy
criteria for aspects targeting the faults presented in the previous
section.

Due to the mixed nature of an aspect definition, which can address
both static and dynamic crosscutting using pointcuts, intertype
declarations, and advice, it is not so easy to obtain a single
criterion that allows us to make meaningful statements of the form
``we have tested 75\% of this aspect''. Instead, we will define
different criteria for the various elements in an aspect definition.

Introducing a new method $m$ in a class $C$ is akin to directly adding
the method to $C$.  Therefore, normal coverage goals such as statement
or branch coverage apply.  However, as we have seen in the fault
model, the most powerful and dangerous introductions are those where
polymorphic methods are added.  Therefore, adequacy criteria
explicitly based on exercising all possible polymorphic bindings are
in place as well.  Rountev \emph{et al} \cite{rountev:2004} include an
up to date overview of criteria for polymorphic bindings.  They
distinguish the \emph{all-receiver-classes} criterion which requires
exercising all possible classes of the receiver object at a call site,
and the \emph{all-target-methods} criterion which requires exercising
all possible bindings between a call site and the methods that may be
invoked by that site.

The intertype declaration of a new supertype or interface for a given
class changes the inheritance hierarchy it belongs to.  This, again,
calls for adequacy criteria taking polymorphic calls into account.
Observe that these adequacy criteria take into account all \emph{call
  sites} within the rest of the application.  Thus, polymorphic
coverage goals are not just a percentage of the aspect definition
itself, but a percentage of how well affected call sites are covered.

To deal with adequacy for pointcuts, we will say that a test case $T$
exercises a pointcut $P$ if $T$ activates advice at a join point
captured by $P$.  An adequate test suite for a pointcut should
maximize our chance of finding errors in the pointcut.  We distinguish
primitive pointcuts and compound pointcuts built from conditional
operators.

Primitive pointcut operators (such as \texttt{call}, \texttt{cflow},
and so on) can capture a multitude of join points.  Which of these
should we ensure we execute in order to maximize our chance of finding
errors in the pointcut?  In most cases we cannot answer this question,
so an arbitrary join point will do.  For signature or type matching
involving wild cards, we can arrive at the equivalent of traditional
boundary testing by ensuring we have one case where the asterisk
matches the empty string, and one where its match is non-empty.  When
matching types in a hierarchy, for example in a \texttt{call(*
  Class+.*)} expression, the class named is a boundary.  Using the
one~$\times$~one criterion \cite{Binder:2000} insisting on one point
on the boundary, and one just outside it, we would obtain one test
case for \texttt{Class}, and one for each immediate superclass
(interface) above it in the hierarchy.

Tests for pointcut expressions composed from multiple conditions
should exercise every relevant condition combination.  In traditional
testing, the most rigorous approach is to test each true/false
combination, leading to $2^N$ test cases for an expression with $N$
conditions.  Alternatively, the \emph{Each-Condition/All-Conditions}
criterion can be used which leads to $N+1$ test cases by insisting on
one test case for each condition making that condition true and all
others false, in addition to one making all conditions true (for
\emph{and} logic, replacing true and false for \emph{or} logic)
\cite{Binder:2000}.  Pointcut logic, however, is different from normal
Boolean logic, for example in that certain operators (such as
\texttt{target}) are primarily meaningful in combinations with others.
Moreover, there are typical idioms for using pointcuts, such as a
sequence of a general pointcut (such as all public calls) conjuncted
with several exceptions (each using the negation operator) for classes
or methods that are to be excluded from the pointcut.

Test adequacy for \emph{advice} itself can again be based on branch or
statement coverage.  It is most natural to compare advice with a
method that is called at relevant join points.  Thus, to achieve branch
coverage for the advice, we do not need to exercise all branches at
every join point: it suffices to find one join point at which we
exercise all branches.

Furthermore, it is natural to insist that each join point at which the
advice is activated is exercised.  Typically, a test suite achieving
statement coverage for the full unwoven application will get a far way
in covering all join points.  One may be tempted to think that covering
all captured join points also achieves adequate pointcut coverage.
This, however, is not the case, since the pointcut may be defined as a
complex expression, parts of which are used to \emph{prevent} firing
at a particular join point.

Last but not least, there may be (abstract) reusable aspects whose
pointcuts do not refer to particular (named) classes or methods they
should be woven into.  To test such aspects, a stub application needs
to be created, to which the aspect can be applied.  When creating
these stubs, the test adequacy issues presented above can be used as a
guideline, ensuring for example that it is indeed possible to exercise
all conditions in the pointcut.  As far as we can see, reusable
aspects themselves provide no further test adequacy constraints.

\subsection{The \bettar Testing Strategy}

The test strategy combining the fault model and various adequacy
criteria consists of three steps:
\begin{description}
\item[Responsibility-Based Testing (Black Box):] Create or identify a functional
  test suite for the concern at hand.  Focus on answering the question
  whether the implementation of the concern does what it is supposed
  to do.
  
\item[Risk-Based Testing (Grey Box):] Use the fault model to refine the test
  suite so that faults due to the refactoring process as well as the
  (aspect-oriented) target solution are most likely to be captured.
  
\item[Source-Based Test Adequacy Validation (White Box):] 
  Inspect the coverage of
  the test suite developed so far (either by running it on instrumented
  code or by manual analysis), and verify that relevant test
  adequacy criteria are indeed met.  If not, return to the previous
  steps to create additional responsibility- or risk-based test cases
  until the adequacy criteria are fullfilled.
\end{description}  

If the refactorings do not affect the external interfaces of the
classes under test, the test suite can be applied to both the original
and the refactored system.  This has the advantage that one can be
certain that each new test also successfully passes on the old
implementation.  If refactoring \emph{does} require making adaptations
to the test suite, the following approaches are possible:
\begin{itemize}
\item Refactor the test suite so that it exercises more global
  functionality instead of invoking the modified methods directly,
  making it more robust to future implementation changes but
  potentially making it harder to achieve the desired coverage;
\item Apply an additional refactoring to the application offering for
  example an additional interface abstracting away from implementation
  differences between the original and target solution;
\item As a last resort, we could give up on our attempt to apply the
  test suite to the original system, and apply new tests to the new
  system only, thereby losing them as safeguard against behavior
  modification during refactoring. In our case study we were never
  forced to do this, and could always refactor the test suite to permit
  testing of both versions.
\end{itemize}

%%% Local Variables: 
%%% mode: latex
%%% TeX-master: t
%%% End: 

\section{\ajhd: An Open Source Aspect-Oriented Showcase}
\label{Sec:ajhd}

To experiment with the feasibility of adopting aspect-oriented
solutions in existing software and demonstrate the strategy proposed
earlier, we have created \ajhd: an aspect-oriented refactoring of
\jhotdraw,\footnote{\url{jhotdraw.org}, version 5.4b1} a relatively
large and well-designed open source Java application.  In order to
allow other researchers to benefit from our work and to enable
comparative software evolution research on a real-life aspect-oriented
system, we decided to release \ajhd as an open source
project.\footnote{\url{ajhotdraw.sourceforge.net} \emph{Note for the
    reviewer: we are currently in the process of cleaning up our
    refactored code, upgrading it to JHotDraw 6.0, and moving it to
    the sourceforge server (see the release plan on sourceforge).  Our
    internal version is available upon request.}}

The next sections give a description of the case study and motivate
the choice for both the application and the refactoring language.

\subsection{The \jhotdraw Drawing Framework}

\jhotdraw is a (GUI-based) framework for drawing technical and
structured 2D graphics.  The application was originally developed as
an exercise to show a good use of object oriented design patterns in a
Java implementation. The fact that \jhotdraw is considered a
well-designed application makes it an ideal candidate for
aspect-oriented migration as it is unlikely that evolvability
improvements can be made otherwise.
The version of \jhotdraw analyzed consists of approximately 
40,000 lines of code, 300 classes, and  2800 methods.

The \jhotdraw editor comprises drawing tools, a set of user defined
(geometrical, image, text, etc.) figures, drawing views, and a
collection of (tool and menu-associated) commands.  A number of
additionally supported features include (re-)storing drawings
(from/)to storage devices, undo/redo activities for commands, and
animation functions.

\subsection{Evolving a \jhotdraw Test Suite}

The version of \jhotdraw  under study was shipped without a
test suite. At the time of writing, the most recent version
(v6.0beta1) has a number of \emph{empty} test classes, automatically
generated using a Java doclet. The intent is to fill them with test
cases, but so far these have not been made available.\footnote{We have
  agreed with the \jhotdraw maintainers that our test suite will be
  integrated in their project.

}

To safeguard our refactorings, we have developed our own functional
test cases on a by need basis. Since we were, initially, not familiar
with the \jhotdraw code, the development of these test cases served as
our program comprehension strategy (\cite{Deu01.xpu}): We formulated
hypotheses on \jhotdraw's implementation, expressed them as test
cases, and then attempted to refute them by running these test cases.
The test suite that was developed passes on the original, pure object
oriented version of \jhotdraw, as well as on \ajhd, the refactored,
aspect-oriented version.

Our test suite is based on the JUnit framework~\cite{BG98}. Moreover,
where needed, we make use of Java 1.4 assertions to ensure that the
alterations did not break invariants or pre- and postconditions. Since
we did wanted to minimize the number of changes to the \jhotdraw code, we injected these
assertions by means of aspects.  For ensuring invariants, the aspect
contains an inter-type declaration giving relevant classes a boolean
\texttt{invariant} method, as well as a pointcut ensuring this method
is indeed checked before and after each public method.
Observe that the assertion aspects are independent of our test suite,
and can be woven into the production version of \jhotdraw as well
in order to simplify debugging.

\subsection{\ajhd Organization}

\ajhd is organized into two parts: (1) the main project is the
\aspectj{} implementation of the system, where the identified
crosscutting concerns are refactored to aspects; (2) the test
subproject (JHDTest) comprises all the test cases aimed at ensuring
equivalence between the the original Java solution and the refactored
\aspectj{} one.
The aspects are put in separate packages, one per concern.
Changes to the original files are restricted just to removing concerns
that have been migrated to aspects.

The tests suite can be compiled with and executed on the archived binary
files (jar) of any of the two solutions.
Building and executing 
the test suite is automated using \textsc{Ant}.\footnote{\url{ant.apache.org}}

%%% Local Variables: 
%%% mode: latex
%%% TeX-master: "teststrategy"
%%% End: 

\section{Refactoring of Selected Concerns from \jhotdraw}
\label{Sec:selectedrefacts}

Previously, we have employed fan-in analysis for the identification of
crosscutting concerns in \jhotdraw~\cite{FanIn04}.  This resulted in
10 types of concerns that were candidates for refactoring into an
aspect.  In this section we discuss three of these concerns
(persistence, contract enforcement, and undo) in considerable detail,
covering the \bettar steps aspect design, refactoring design, and test
suite design.

A transparent, gradual process of refactoring is important for
building confidence in the aspect-oriented solution. Therefore, our
refactorings aim at maintaining the conceptual integrity and stay
close to the original design. An additional advantage of this approach
is that this preserves the understandability of the refactored system
for the original maintainers.

\subsection{Refactoring The Persistence Concern}

\subsubsection{Aspect Design}

Drawings in \jhotdraw are collections of figures that can optionally
be stored and recovered (write/read operations) by the application.
The concern denoting this functionality, persistence, is defined by
the \emph{Storable} interface that declares two methods,
\emph{write(StorableOutput)} and \emph{read(StorableInput)}.
The entire hierarchy of storable elements in a drawing comprises 94
interfaces and classes, of which 40 belong to the \emph{Figure} class
hierarchy.

Because the persistence concern is already distinguished in the
original design, refactoring it to an aspect is fairly
straightforward.  The aspect can use \emph{introductions} in order to
have the persistent elements of a drawing (e.g., figures) implement
the \emph{Storable} interface.  If not all variables comprising the
state of the class are accessible through public getters and setters,
the aspect will need access to private members as well.  The \aspectj{}
way to achieve this is by declaring the aspect \emph{privileged}.

The implementation of the \emph{Storable} interface also implies an
interesting enforcement constraint: ``Objects that implement this
interface and that are resurrected by StorableInput have to provide a
default constructor with no arguments.''  This constraint cannot be
enforced by \aspectj{}.  A similar situation occurs when refactoring
bean objects (see the ``Bean Aspect'' example in \cite{AspectJPG})
that must define no-argument constructors.

A concern related to persistence is \emph{serialization}, which in
\jhotdraw is also implemented for the \emph{Figure} hierarchy.  According to
Java API specification, classes requiring special handling during
de-serialization, such as a number of figures in \jhotdraw, must
implement a special private method
(\emph{readObject(ObjectInputStream)}).  \aspectj{} does not support
introduction of private members into target classes.  The visibility
of the inter-type declarations relates to the aspect and not to the
target class.  Although already acknowledged as a shortcoming (see
\cite{AspectJPG}) the language interpretation of visibility prevents a
consistent refactoring of similar kinds (persistence and
serialization) of crosscutting concerns.

A summary of the various issues is provided in
Figure~\ref{Fig:Persistence}.

\subsubsection{Refactoring Design}
The refactoring itself is fairly straightforward, and just consists of
moving read and write method implementations to the persistence
aspect.  The complete refactoring of the persistence concern can
generally be described as \emph{Extract Interface Implementation}
as discussed by \cite{laddad2003:refactoring}.

\begin{figure}
 \leavevmode\centering
  \begin{tabular}{|>{\bf}p{.19\hsize}p{.78\hsize}|}
    \hline
   Old situation & 
   Objects requiring persistence implement the \emph{Storable} interface.
   \\ 
    Aspect solution &  
    Implementation of \emph{Storable} interface moved to aspect by means
      of introductions.  
   \\
   Code size &  Remains the same
   \\
   Benefits &
   All persistence related code in one aspect; classes oblivious of whether
   they can be made persistent.
    \\
  Risks &
   Encapsulation broken since persistence aspect requires privileged access.
  \\
  \aspectj{} issues &
  Zero-argument constructor cannot be enforced; 
  Private methods cannot be introduced.
  \\ \hline
 \end{tabular}
 \caption{Refactoring of the Persistence Concern\label{Fig:Persistence}}
\end{figure}

\subsubsection{Test Suite Design}

Testing the persistence aspect is relatively simple.  We nevertheless
discuss it in some detail, since the way of testing can be reused for
other more complicated concerns that we will discuss next.

When refactoring persistence to an aspect we run a number of risks:
The first is that in our aspect, we accidentally introduce a read or
write method body for a given figure in the wrong class.
The second is that we make an error when copy-pasting the body of 
a method to an aspect.
Last but not least, our removal of the persistence code from, e.g.,
figures may be incomplete.

In order to test persistence we proceed as follows: First, we create a
top level \emph{StorableTest} class, which has a test method that (1)
creates a \emph{Storable} (typically a figure), (2) writes it to a
stream, (3) reads it back into a different object, and (4) checks the
equivalence between the two. Next, the creation of the actual figure
is deferred to subclasses of the \emph{StorableTest} class using a
virtual factory method.  Thus, the test hierarchy mimics the hierarchy
of classes to be stored. Finally, our equivalence checking method
should be based on structure, not on object identity. Such a method is
not included in the \jhotdraw implementation.  We injected this method
into the class hierarchy using an aspect. Observe that a collection of
static equivalence methods included in, for example the test class,
would not work, since the equivalence method must be polymorphic --
which can be achieved by means of introductions in an aspect but not
by means of static methods.

This strategy implements Binder's \emph{Polymorphic Server Test} test
design pattern~\cite{Binder:2000}.  It can be used to verify that
subclasses conform with superclass behavior, and that we are setting
up a correct polymorphic hierarchy.  It requires exercising each
superclass test case to every possible subclass.  In other words, we
can reuse the write-read-compare test case for every subclass of
\emph{Storable}.

\subsection{Contract Enforcement in Commands}

\subsubsection{Aspect Design}

\jhotdraw makes use of the Command design pattern in order
to separate the user interface from the underlying model,
and in order to support such features as undoing and redoing
user commands.
Each command has to realize the \emph{Command} interface, for which
a default implementation is provided in the \emph{AbstractCommand}
class.
The key method is \emph{execute}, which takes care of actually carrying
out the command (such as pasting text, inserting an image, etc.).

Each \emph{execute} method should start with a consistency check 
verifying that the underlying ``view'' exists. 
Therefore, each concrete implementation of \emph{execute} starts with 
a call to the \emph{execute} implementation in the superclass, which is always
the one from the \emph{AbstractCommand}.
This is illustrated in Figure~\ref{Fig:SuperExecute}.

\begin{figure}
\leavevmode\centering
{\scriptsize
\begin{boxedverbatim}

public class AbstractCommand implements Command {
  ...
  public void execute() {
    if (view() == null) {
      throw new JHotDrawRuntimeException(
        "execute should NOT be getting called when view() == null");
} } }

public class PasteCommand extends AbtractCommand {
  ...
  public void execute() {
    super.execute();
    ...
} }

\end{boxedverbatim}
}
\caption{Contract Enforcement using a super method idiom.\label{Fig:SuperExecute}}
\end{figure}

This is a typical example of what is called ``contract enforcement''
in the \aspectj{} manual \cite{AspectJPG}.
We implemented it using a pointcut capturing all \emph{execute} methods,
putting the check itself in the advice.
Observe that mimicking the implementation where the check is in a
super method is not possible in \aspectj{}:
super methods cannot be accessed when advising a method.
The resulting solution is shown in Figure~\ref{Fig:AspectExecute}.

\begin{figure}
\leavevmode\centering
{\scriptsize
\begin{boxedverbatim}

pointcut commandExecute(AbstractCommand aCommand) :
  this(aCommand)
  && execution(void AbstractCommand.execute())
  && !within(*..DrawApplication.*);

before(AbstractCommand aCommand) : commandExecute(aCommand) {
  if (aCommand.view() == null) {
    throw new JHotDrawRuntimeException("...");
} }

\end{boxedverbatim}
}
\caption{Enforcing the consistency check using before advice.\label{Fig:AspectExecute}}
\end{figure}

The only surprise in this figure may be the \texttt{within} clause in the pointcut.
It turns out that \emph{anonymous} subclasses of \emph{AbstractCommand} do
not implement the consistency check. Such classes are used for simple 
commands such as printing, saving, and exiting the application.
Since \aspectj{} does not provide a direct way to exclude anonymous classes
in a pointcut, we used the \texttt{within} operator to exclude executions 
occurring in the context of the top level object creating the full user interface.
One can also argue that the anonymous classes should include the check
(in which case the exclusion can be omitted from the pointcut),
but at present we focus on keeping the behavior as it was, not on
modifying it.

The main benefit of the aspect approach is that consistency checks
cannot be forgotten.
This is illustrated by the anonymous classes, but also by one
non-anoymous command,\footnote{Namely, the \emph{UndoableCommand}.}
which does not extend the
\emph{AbstractCommand} default implementation.
Consequently, it cannot reuse the consistency check using a supercall.
Inspection of the \emph{execute} implementation, however, clearly shows that the
code exits with a null pointer exception in case the check fails.
This suggests that the aspect that we are looking for should implement
the check not only for the \emph{AbstractCommand} class, but
for all implementations of the \emph{Command} interface.
Again, our current implementation does not yet
do this, but only injects the implementation in subclasses of
\emph{AbstractCommand}.

A summary of the main issues in the Contract Enforcement refactoring 
is provided in Figure~\ref{Fig:ContractEnforcement}.

\begin{figure}
   \leavevmode\centering
   \begin{tabular}{|>{\bf}p{.19\hsize}p{.78\hsize}|} \hline
   Old situation & 
   Each concrete \emph{execute} invokes its super \emph{execute}
   in order to conduct certain consistency checks.
   \\ 
    Aspect solution &  
    The consistency check is implemented as advice,
    which is invoked before each call to \emph{execute},
    as captured in a simple pointcut.
   \\
   Code size &  
   17 explicit consistency calls replaced by one pointcut;
   consistency check itself moved from class to advice.
   \\
   Benefits &
   Reliability: it becomes impossible to forget the consistency check.
   Omitted checks can be fixed automatically thanks to the refactoring.
    \\
  Risks & 
   Check required that omissions are not on purpose.
  \\
  \aspectj{} issues &
  No direct support to capture anonymous classes;
  Cannot refer to super methods in method advice.
  \\ \hline
 \end{tabular}
 \caption{Refactoring Contract Enforcement for Commands.\label{Fig:ContractEnforcement}}
\end{figure}

\subsubsection{Refactoring Design}
The restructuring can generally be described as an \emph{Advise Method Overrides}
refactoring, as presented in Section~\ref{Sec:catalogrefacts}.

\subsubsection{Test Suite Design}

Simple as the pointcut in Figure~\ref{Fig:AspectExecute} may be, it
nevertheless illustrates some of the issues involved in testing
refactorings that make use of pointcuts.

First of all, adequate testing of the consistency check in the
original (non-aspect) \jhotdraw version would typically correspond to
branch coverage.  This yields two test cases for the top level execute
method (one in which the consistency check passes, and one in which it
fails) in addition to one dedicated test for the \emph{execute}
implementation in each subclass.  Since the super call can be resolved
statically, even polymorphic adequacy models will not add test cases
to this.

It is interesting to observe that such a test suite would not capture
the subtleties involved in designing the aspect from
Figure~\ref{Fig:AspectExecute}.  For example, the test suite does not
exercise anonymous classes, nor \emph{execute} methods occurring
outside the scope of \emph{AbstractCommand}.

The aspect-specific test adequacy criteria as discussed in
Section~\ref{Sec:teststrategy}, however, do suggest creating the
relevant additional test cases.  Inspection of the pointcut leads to
the following tests:
\begin{itemize}
\item Since \emph{AbstractCommand} occurs in a type match, we would
  like to test classes just off this boundary as well, leading to a
  test case checking what happens for the \emph{Command} interface
  itself.
\item Since the pointcut is a conditional expression, we also want to
  investigate what happens if one of the conditions fails.  This means
  that we want to verify that the \texttt{within} clause does fire for
  anonymous classes.
\end{itemize}

Actually creating these test cases may, however, not be as easy as it
seems.  Testability is affected by controllability and observability,
which are poor for anonymous classes and join point execution.

In order to verify (observe) that our pointcut from
Figure~\ref{Fig:AspectExecute} does indeed capture anonymous classes
correctly, we created special advice used for testing purposes only,
which keeps track where a certain pointcut expression has fired.  To
do this, we first refactored the aspect so that the individual
conditions are in separate pointcuts, as shown in
Figure~\ref{Fig:ExecutePointcut}.  The production aspect uses these
pointcuts to perform the consistency check at the right places.  The
testing aspect uses exactly the same pointcut definitions to weave in
code that keeps track of where (i.e. at which joinpoints) those
pointcuts have fired. This set of joinpoints is then used to verify
intended behavior.

\begin{figure}
\leavevmode\centering
{\scriptsize
\begin{boxedverbatim}

abstract aspect ContractEnforcementPointcut {

  pointcut commandExecute(AbstractCommand aCommand) :
    this(aCommand)  
    && inExecuteMethod() 
    && ! inAbstractClass()

  pointcut inAbstractClass() :
    within(*..DrawApplication.*);

  pointcut inExecuteMethod() {
    execution(void AbstractCommand.execute());

} }

\end{boxedverbatim}
}
\caption{Separate pointcuts for each condition to
  improve aspect testability.\label{Fig:ExecutePointcut}}
\end{figure}

Concerning controllability, the instances of the anonymous classes are hard to access.
They are normally activated via a mouse event, which must be mimicked
in order to trigger the command's \emph{execute} method.
We avoided the need for generating mouse events by using an aspect:
we intercept the constructors for anonymous command classes,
and collect them in a set: after the full application has been built
we can apply the execute method to each command.

\subsection{Refactoring the Undo Concern}

\subsubsection{Background and Current Approach}

Support for ``undo'' is a newly added feature in the analyzed version
of \jhotdraw.  As can be imagined, it is a concern that crosscuts
across many different classes.  More than 30 elements of the \jhotdraw
framework, comprising \emph{commands}, \emph{tools} and
\emph{handles}, have associated undo constructs to revert the changes
spawned by their underlying activities.  The discussion here will
focus on the \emph{commands} group, as it is the largest in terms of
defined undo activities.

\begin{figure}
  \leavevmode\centering \includegraphics[width=.75\hsize]{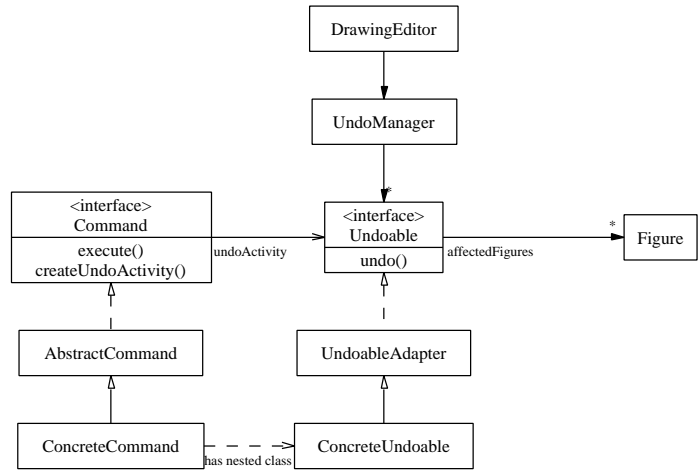}
  \caption{\label{Fig:UndoParticipants}%
    Participants in \jhotdraw's undo implementation.}
\end{figure}

Some participants in \jhotdraw's undo implementation are
shown in Figure~\ref{Fig:UndoParticipants}:
\begin{itemize}
\item Each command is associated with one \emph{undo activity}, whose
  method \emph{undo} can be invoked to revert the command.
\item The undo activity is implemented in a nested class of the
  command, which is instantiated using a factory method called
  \emph{createUndoActivity}.
\item The primary abstraction in the undo activity is the list of
  affected figures: when the command's execute method is invoked, the
  relevant state of the affected figures is stored in the undo
  activity.
\item Undo activities are maintained on a stack by the undo manager.
\end{itemize}

\subsubsection{Aspect Design}

The aspect solution to undo we propose consists of associating an
undo-dedicated aspect to each undo-able command. The aspect implements
the entire undo functionality for the given command, while the
associated class remains oblivious to its secondary concern. By
convention, for enforcing the relation with the command class, each
aspect will consistently be named by appending ``UndoActivity'' to the
name of the command class.  In a successive step, the command's nested
\emph{UndoActivity} class moves to the aspect.
The factory methods for the undo activities
(\emph{createUndoActivity()}) also move to the aspect, from where they
are introduced back into the associated command classes using
inter-type declarations.

The statements in the \emph{execute} method that are responsible for setting up the 
undo activity, are taken out of the \emph{execute} method, and woven into it by means of advice.
In some cases the corresponding pointcut simply needs to capture all execute method calls;
in other cases the pointcut is more complex, depending on the way the
undo code is mixed with the regular code.

\begin{figure}[t]
\leavevmode\centering
{\scriptsize
\begin{boxedverbatim}
public class PasteCommand extends FigureTransferCommand {
    public void execute() {
        ...
        FigureSelection selection = 
            (FigureSelection)Clipboard.getClipboard().getContents();
        if (selection != null) {
            setUndoActivity(createUndoActivity());
            ... //core command logic and other undo setup
            FigureEnumeration fe = insertFigures(...);
            getUndoActivity().setAffectedFigures(fe);
            ...  
}   }   }
\end{boxedverbatim}
}
\caption{\label{Fig:OriginalPasteCommand}%
The original PasteCommand class.}
\end{figure}

As an example, consider the paste command, whose execute method consists of
retrieving the selected figures from the clipboard, inserting them into the current view,
and clearing the clipboard.
All this is done in a single method, using local variables and if-then-else statements
to deal with such situations as an empty clipboard.
The undo aspect will require the same conditional logic, and access to the same data
in the same order.
The following aspect solutions are possible:
\begin{itemize}
\item
If all getters are side effect free, an approach is to setup the undo activity in a simple
before advice.
In \jhotdraw, however, this is not the case, for example because of figure enumerators that
have an internal state.
\item
The alternative route is to intercept relevant getters, keep track of the data locally in the advice
as well, and inject advice after all data has been collected.
This is the approach we follow, but some of the pointcuts are somewhat artificial. Figure~\ref{Fig:UndoAspectJHD}
illustrates such a pointcut in the undo aspect for the \emph{PasteCommand}, which is also shown in 
figure~\ref{Fig:OriginalPasteCommand}. The \emph{execute\_callClipboardgetContents()} pointcut captures the call
that sets the reference to be checked by both the command's core logic and the undo functionality in the aspect.
\item
The last possibility is to refactor the long \emph{execute} method into smaller steps
using non-private methods.
The extra method calls can be intercepted allowing smooth extension with setting up the
undo activity, at the cost of creating a larger interface and breaking encapsulation.
\end{itemize}

\begin{figure}
\leavevmode\centering
{\scriptsize
\begin{boxedverbatim}
public aspect PasteCommandUndoActivity {
    //store the Clipboard's contents - common condition
    FigureSelection selection;

    pointcut execute_callClipboardgetContents() :
        call(Object Clipboard.getContents()) && withincode(void PasteCommand.execute());

    after() returning(Object select) : execute_callClipboardgetContents() {
            selection = (FigureSelection)select;
    }
    ... 
        
    pointcut executePasteCommand(PasteCommand cmd) :
        this(cmd) && execution(void PasteCommand.execute());
        
    // Execute undo setup
    void after(PasteCommand cmd) : executePasteCommand(cmd) {
        // the same condition as in the advised method
        if(selection != null) {
            cmd.setUndoActivity(cmd.createUndoActivity());
            ...
            cmd.getUndoActivity().setAffectedFigures(...);
}   }   }
\end{boxedverbatim}
}
\caption{\label{Fig:UndoAspectJHD}%
The undo aspect for PasteCommand.}
\end{figure}

The resulting system differs in two ways from the original design.
First, the original design uses static nested classes 
to enforce a syntactical relation between the undo
activity and its enclosing command class. Since the \aspectj{} 
mechanisms do not allow introduction of nested classes, the post-refactoring association will 
only be an indirect one, based on naming conventions. This is a weaker connection than the one
provided by the original solution. 
A second difference is that the visibility of certain methods has been altered,
since \aspectj{} cannot be used to introduce, for example, the required factory method as
\emph{protected}.

\subsubsection{Refactoring Design}
The complexity of the refactoring is determined by the complexity of unplugging
undo from the commands themselves.
We distinguish different levels of unpluggability:
\begin{enumerate}
\item 
  The nested undo activity class of the command, and all its uses can be safely removed from the command.
  The fairly simple \emph{ChangeAttributeCommand} class is an example in this category.
\item
  The command's core logic makes use of some of the data stored in the undo activity.
  This is typically done for the list of affected figures.
  Since there is no real need for this, we could easily refactor the core logic so that
  it does not refer to the undo activity anymore.
\item
  The nested undo activity not only deals with undo, but also contains core logic needed for the 
  proper execution of the command. An example is the \emph{InsertImageCommand}:
  its undo activity contains a method called \emph{insertImage} which actually inserts the image
  (instead of undoing it).
  We consider this a design violation.

  Our solution consists of applying traditional refactorings before starting with the
  aspect refactoring, so that the command does not depend on the undo activity anymore.
\item
  The nested undo activity is not only used for this particular command, but also for
  similar commands. This is the case for the \emph{PasteCommand}.
  Our aspect refactoring will rename the undo activity, and hence requires a simple change
  to these commands.
\end{enumerate}

We anticipate that any non-trivial aspect refactoring will require
similar object-oriented refactorings, before the crosscutting concern
can be taken out of the available system.  A more detailed discussion
of the undo concern refactoring, accompanied by code snippets, is
presented in~\cite{UndoWare04}.

\begin{figure}
 \leavevmode\centering
   \begin{tabular}{|>{\bf}p{.19\hsize}p{.78\hsize}|}
    \hline
   Old situation & 
   Each command's \emph{execute} sets up a corresponding 
   undo activity, which is implemented through a nested class.
   \\ 
    Aspect solution &  
    One aspect per command, which contains the undo activity implementation,
    and introduces the association into the command.
    Execute method intercepted to setup the proper undo activity state.
   \\
   Code size &  
   Remains the same.
   \\
   Benefits &
   Strong tangling between commands and their undo activity eliminated;
   commands are easier to understand.
    \\
   Risks & 
   Undo activity may require sophisticated pointcuts to intercept
   all relevant state modifications of the command;
   Refactoring of commands needed in order to unplug undo support from them.
  \\
  \aspectj{} issues &
  No support for introducing nested classes.
  Visibility affected since protected methods cannot be introduced.
  Modular reasoning affected by keeping track of data set in the advised method.
  \\ \hline
 \end{tabular}
 \caption{Refactoring Undo.\label{Fig:UndoRefactoring}}
\end{figure}

\subsubsection{Test Suite Design}

In testing undo, we essentially combine the testing approaches
of the persistence and contract enforcement concerns discussed
previously.

First of all, we create a reusable test suite at the \emph{Command}
level. This test can be used for any command subclass,
and ensures that each subclass complies with then intended semantics.
This test set takes care of:
\begin{itemize}
\item 
  Setting up an appropriate \jhotdraw application in which a 
  concrete command can be created. The actual command created is
  deferred to subclasses of the test class.
\item
  Bringing the application in a setting in which the execute
  can be carried out (for example, many commands require that
  some figures in the drawing are selected), and 
  actually invoking it.
\item
  Comparing the effects of the command execution with the
  intended behavior --- this step is specific to the actual command
  and deferred to subclasses. It usually consists of comparing the
  modified selected figures with a set of figures actually constructed
  in the test case. 
\item
  Invoking the \emph{undo} method on the command's
  undo-activity, and comparing that the effects are indeed canceled.
  Again, this comparison typically involves the set of affected
  figures.
\end{itemize}
Thus, the test case follows the template method design pattern,
and defers the details of certain steps to its subclasses.

To test the various pointcuts, the approach described for
contract enforcement was adopted, weaving in special advice
that allowed us to observe which pointcut actually fires.

%%% Local Variables: 
%%% mode: latex
%%% TeX-master: t
%%% End: 

\section{Contributing to the Catalog of Refactorings}
\label{Sec:catalogrefacts}

Several authors have proposed catalogs of aspect-oriented refactorings
\cite{laddad2003:refactoring,MPM04,Monteiro03}, in the spirit of
Fowler's catalog of object-oriented refactorings \cite{fowler1999}.
We were able to reuse several of these existing refactorings, such as
Monteiro's \emph{Encapsulate Implements with Declare Parents}, and
\emph{Move Method from Class to Inter-type}, or Laddad's \emph{Extract
  Method Calls} refactoring which encapsulates calls to a method from
multiple places into an aspect.

In this section we add our contribution to these existing catalogs,
casting some of the experiences we obtained from building \ajhd into
generally reusable refactorings.

An open question is at what level of abstraction such refactorings
should be defined.  Is introducing some design pattern considered a
refactoring?  It is, but Fowler's book has explicit refactorings
described for just a few design patterns, not all.  The reason for
this is, most likely, that the mechanics for introducing such a design
pattern can hardly be described in a reusable way, and for that
reason the refactoring description would not add much useful
information to the pattern description.  In this respect an
interesting approach is taken by Kerievsky \cite{Ker04}, who
explicitly addresses refactorings to patterns.  He focuses on a subset
of the design patterns, namely those for which common coding tricks
are known that do not yet provide the benefits of using the full
pattern, such as in his \emph{Replace Hard-Coded Notifications with
  Observer} refactoring.

A similar distinction holds for aspect refactorings.  Introducing
refactorings for each of the prototypical concerns listed in, for
example, the \aspectj programming guide \cite{AspectJPG} may not be
particularly useful.  But in some cases, the ``old'', non-aspect
solution can be reasonably well described (for example an Observer
implementation following the guidelines from \cite{Gamma1994}), and it
does make sense to describe how such an implementation can be
refactored into an aspect solution (such as the one from
\cite{DPAOP02}).

If we look at the refactorings from Monteiro, these can be categorized
as fairly technical, elementary refactorings, such as introducing an
inter-type declaration \cite{MPM04}.  The refactorings from Laddad
\cite{laddad2003:refactoring} are more of a mixed style, some being
elementary, others being closer to typical concerns from the \aspectj
manual.  Below we try to provide some building blocks for creating
refactoring descriptions that give concrete advice how certain
concerns can be turned into aspects.

\subsubsection{Move Role to Aspect}
Though not discussed in the previous section, several of our
refactorings involve the creation of an aspect-oriented implementation
of a design pattern. As an example, \jhotdraw contains several
instantiations of the \emph{Observer} pattern, which we essentially
implemented according to the approach proposed by Hannemann and
Kiczales \cite{DPAOP02}.

The participants in this pattern can be an \emph{observer} or a
\emph{subject}.  The existing \jhotdraw implementation does have a
separate interface for the observer role, but not for the subject
role.  We propose to refactor this and introduce a subject interface
via an aspect in order to: (a) make the two different roles explicit,
and (b) remove the observer pattern details from the primary concerns.
Note that in some cases, one class can be involved in multiple design
patterns adopting different roles for them.  For example, a composite
figure is a subject as well as an observer, listening to changes in
its subfigures while being listened to by, for example, drawings.  The
total number of methods implemented by such multi-role classes can be
substantial, making them hard to understand; a problem addressed by
moving the roles to aspects.

Thus, \emph{Move Role to Aspect} creates an interface for a particular
role in a design pattern, and superimposes this role on an existing
class by means of an aspect.

\subsubsection{Move Observer to Aspect}
A more high level refactoring is to move an observer implementation
into an aspect. This is a compound refactoring, involving three
elementary steps: first, the \emph{Move Role to Aspect} refactoring is
applied twice, once for the subject and once for the observer role.
Subsequently, the calls made in subjects to notify the observers of
changes are captured into a pointcut and extracted into advice.

\subsubsection{Override Method with Advice for Overlapping Roles}
\begin{sloppypar}
  Just like one class can fullfil multiple roles from one or more
  different design patterns, one method can implement features related
  to multiple roles.  This is common in \emph{Java Swing} design and
  also occurs in one of our \jhotdraw refactorings.  This refactoring
  dealt with the \emph{CommandMenu}, which acts as both \emph{view}
  and \emph{controller} for the interactive drawing editor of the
  application.
  The method exhibiting the overlapping roles, \emph{checkEnabled()},
  enables/disables menu items according to the status
  (executable/non-executable) of the command to be activated when the
  item is selected.  Although the method belongs to the interface of
  the \emph{view} component, allowing to set the view's elements
  status, its implementation relies on \emph{controller} decisions.
\end{sloppypar}

The proposed refactoring places the method's definition into the
interface for the role to which it belongs, in this case, the
\emph{view} role, making it accessible to the developer of the GUI.
Furthermore, the controller aspect uses an around advice to override
the default behavior of the method and to make it
context(command)-aware.

\subsubsection{Advise Method Overrides}
This refactoring aims at removing duplication arising from statements
that are common to (the start or end of)
all method overrides of a given (superclass) method.
Such statements are replaced by advice to any refinement of the superclass method.
Examples in \jhotdraw include the contract enforcement we discussed
previously (the check at the beginning of each \emph{execute} method),
as well as a call to the \emph{checkDamage} method that is contained
at the end of each \emph{execute} method.
%%% Local Variables: 
%%% mode: latex
%%% TeX-master: t
%%% End: 

\section{Discussion}
\label{Sec:discussion}

What did we learn from refactoring \jhotdraw to aspects and 
validating behavior conservation by means of testing?

First of all, we once again learned that testing is actually needed
for such refactorings. In several cases, we detected errors in our
pointcuts, introductions, and copy-paste activities thanks to our test
suite.  Although all of us will agree with this need for testing, it
is alarming, to say the least, that neither the popular textbooks on
aspect-oriented programming (such as
\cite{laddad2003,AspectJPG,maj2003}) nor the existing work on
aspect-oriented refactoring \cite{laddad2003:refactoring,Monteiro03}
provides any advice on how to approach aspect-oriented testing in a
systematic way.

Second, our fault model as well as our adequacy criteria illustrate
how easy it is to make errors during aspect-oriented programs, and how
much needs to be done in order to have a reasonable chance of finding
these errors using tests.  Moreover, both the observability (did this
pointcut fire?)  and the controllability (which inputs will cause a
pointcut to be exercised?)  of aspect-oriented programs typically are
problematic.  Admittedly, at several points in time we were tempted to
omit the testing since it seemed too complicated to create a test
suite capable of achieving the required coverage.  Testing tool
support may very well help here: but this requires an adequacy model
first, which is what we proposed in the paper.

Concerning the refactorings themselves, our experiments illustrate
that being oblvious to future extensions is not as easy as it may
seem.  For example, the undo concern was added only in version 5.4 of
\jhotdraw.  Could this have been implemented as a separate aspect
without modifying \jhotdraw version 5.3?  Our refactoring shows the
direction this would take.  But for some commands, such as the paste
command, artificial pointcuts are needed, which are very brittle if
the underlying primary logic in the command changes.

For most cases, assessing the benefits of an aspect-oriented
refactoring turned out to be a fairly subjective process that is hard
to quantify.  The \emph{aspect design} step looks for such solutions
that would enhance the system's evolvability; that is, to achieve a
better modularization for the, otherwise, scattered and tangled parts
of a concern, and to provide an implementation that better reflects
the concern-based reasoning over the system.  It is not always
apparent, however, in the context of a (relatively) large system as
the analyzed case study, that the new, aspect solution surpasses the
legacy one. Although we argue to have improved the separation of
concerns, for some more complex refactorings, e.g, \emph{undo}, the
downfalls of the aspect-oriented implementation make it difficult to
asses the improvements for the overall system, or even the gains in
modular reasoning over the refactored crosscutting concern.
Difficulties could also occur for less demanding refactorings as for
example, contract enforcement, depending on the uniformity of the
places where the contract needs to be enforced.

Last but not least, it is striking that almost every refactoring we
experimented with raised one or more issues concerning \aspectj (such
as visibility modifiers, nested classes, or anymous classes).  Some of
these limitations are quite technical in nature, and are likely to be
resolved in future versions of \aspectj.  Also, other aspect-oriented
frameworks, such as AspectWerkz~2\footnote{aspectwerkz.codehaus.org} ,
may offer solutions to some of the issues.  Other limitations are more
fundamental (such as the constraint that a class should offer a
zero-argument constructor or the inability to access super methods),
and call for a more rigorous reconsideration of existing
aspect-oriented models.

%%% Local Variables: 
%%% mode: latex
%%% TeX-master: "teststrategy"
%%% End: 

\section{Related Work}
\label{Sec:related}

An important part of research into the area of refactoring to
aspect-orientation has analyzed aspect solutions to a number of
(sometimes complex) concerns that typically crosscut the primary
decomposition of a
system~\cite{laddad2003:refactoring,DPAOP02,laddad2003}.  The
association between the concern and its aspect solution is an
important indication of how a specific language model is intended to
address types of crosscuttings.  However, the specific implications of
applying the refactorings in the context of a large system, where
deviations from the examples used to describe the refactorings are
very likely, are not considered. In this paper we showed some of the
difficulties that arise when these solutions are applied to concerns
in a large system.

A number of authors investigated the possibility of building catalogs
of aspect refactorings.  Monteiro and Fernandes~\cite{Monteiro03}
proposed a set of code transformations from Java to \aspectj{}
specific modularization units, describing steps in a feature
extraction process. The approach has followed the format used by
Fowler~\cite{fowler1999} to describe object-oriented refactorings, and
was further significantly extended~\cite{MPM04}. The study emphasized
the mechanics associated to code transformations as opposed to the
relation with typical crosscutting
concerns~\cite{laddad2003:refactoring,AspectJPG,DPAOP02}.  A similar
list is also proposed by Iwamoto and Zhao~\cite{ZhaoRefact03}, but the
authors do not provide details about any of the specific refactorings.
The attention tends to focus on potential conflicts between the aspect
refactorings and the traditional, object-oriented ones. This issue is
also addressed by Hanenberg \emph{et al}~\cite{Han2003}, as well as
Hannemann \emph{et al}, who discuss the
possibility of a refactoring approach based on a developer-tool
dialog~\cite{murphyrefact03}.

Specific techniques, like program slicing, are employed by Ettinger
and Verbaere~\cite{EttinVerb04} to extract tangled code into method
and further into advice, as an extension of the object-oriented
refactoring to aspects.

Closely related to the work described in this paper, Coady \emph{et
  al} investigate the benefits of aspect-oriented solutions for
evolving operating system code and for better managing its variability
\cite{CK03,BYCF04}. To that end they describe, for example, how the
prefetching concern can be separated from the page handling code in
the FreeBSD kernel code~\cite{CKFS01}.  Although their work aims at
assessing the benefits of aspect-oriented software development, in
contrast to the work presented in this paper, it has not led to a
publicly available aspect-oriented and non-aspect-oriented version of
the same software system which can be used for comparative
experimental software evolution research by other researchers.

However, none of this refactoring work mentions a testing strategy
that accompanies the migration process. The attention given to testing
in the context of aspect-orientation is limited and not with concerns
to refactoring.  Few published test adequacy criteria for
aspect-oriented programming have been formulated: the only work we are
aware of is by Alexander \emph{et al}, who propose a candidate model
and raise a number of research questions~\cite{ABA04}.  Ubayashi and
Tamai~\cite{modelcheck02} use model checking to verify object
crosscutting properties in aspect-oriented programs.  As a first
attempt to define an approach for testing aspect-oriented programs,
Zhao~\cite{ZhaoDFUT} proposes a data-flow-based unit testing. The
tests are oriented towards aspect and class modules that can
potentially be targeted by multiple aspects. Based on the modules'
accessibility three levels of testing are considered, i.e.,
intra-module, inter-module, and intra-aspect or intra-class.

%%% Local Variables: 
%%% mode: latex
%%% TeX-master: t
%%% End: 

\section{Concluding Remarks}
\label{Sec:conc}

Refactoring to aspect-orientation aims at improving the
evolvability and reusability of a system.  Important issues to be
considered in this context are (1) the adequacy of the aspect
solutions discussed by a number of authors when applied to a large
application, (2) the assessment of the support for and improvements
brought by refactoring to aspects, and (3) the challenges of behavior
conservation  when migrating to aspect-supported
implementations.

This paper addresses these problems by proposing a refactoring and
testing strategy to guide the migration process, and successively by
applying it to an open source Java system. The testing strategy aimed
at ensuring migration consistency, introduces an aspect-oriented fault
model and adequacy criteria.  Further, aspect and refactoring designs
are analyzed for selected concerns in the system under investigation,
which also include new, complex examples of crosscuttings.  The
analysis consists of a proposed aspect solution, associated validating
tests, and a trade-off review of the pre- and post-refactoring
implementations.  The difficulties in assessing overall improvements
due to refactoring are turned into considerations about the
suitability of the language features and model for better supporting
the types of identified crosscutting concerns.  We believe that the
development of aspect languages could benefit from catalogs that
associate types of crosscutting concerns to language mechanisms, and
we provide further input for such catalogs.

The paper's main contributions are 
(1) an aspect-oriented fault model and adequacy criteria;
(2) a refactoring strategy that emphasizes testing and the use
    of aspect-oriented solutions;
(3) a detailed discussion of aspect refactorings and their testing implications,
    as carried out on an existing system; and
(4) the initiation of an open source project that can be used
    to experiment with aspect-oriented testing and refactoring,
    and that can be used to compare an object with an aspect solution.

The work described in this paper can be extended in various ways.
First, we will continue to experiment with \ajhd and other case studies,
in order to further extend  the fault model, adequacy criteria, and refactoring catalogs.
Second we will use the proposed models and the experience gained from these
case studies to come up with automated tool support for both testing and
refactoring of aspect-oriented programs.
Last but not least, we will analyze the risks and benefits of the various
aspect solutions, and reflect on ways in which some of the limitations of 
the current solutions can be resolved.

In order to put our work in a broader perspective, we would like to refer to 
Bray~\emph{et al} who state: ``assessment of aspect-oriented software
development in general is still arguably in its early
days''~\cite{BYCF04}. We argue that one of the prerequisites for such
an assessment is the availability of an aspect-oriented and non-aspect
oriented version of the same software system.  Our work aims to create
such versions for a publicly available open source software system and
thereby enables experimental comparative software evolution research
to asses the benefits of aspect-orientation.

\subsubsection{Acknowledgments}
We would like to thank Magiel Bruntink (CWI),
Hylke van Dijk (TU Delft),
Marco Lormans (TU Delft),
and Tom Tourw\'{e} (CWI), 
for reading earlier drafts of this paper.

Partial support was received from SENTERNovem,
(Delft University of Technology, project MOOSE, ITEA 01002,
and CWI, project IDEALS, hosted by the Embedded Systems Institute).

%%% Local Variables: 
%%% mode: latex
%%% TeX-master: t
%%% End: 

\bibliographystyle{plain}
{%\small
\bibliography{main}
}

%% \newpage
%% \setcounter{tocdepth}{2}
%% \tableofcontents

%% \appendix
%% %\input{appendix}
%% \input{scrapbook}

\end{document}